\documentclass[proceedings]{stacs}
\stacsheading{2009}{397--408}{Freiburg}
\firstpageno{397}

\usepackage{color}

\newcommand{\FOR}{\textbf{for }}

\newcommand{\DO}{\textbf{do }}
\newcommand{\ELSE}{\textbf{else }}

\newcommand{\IF}{\textbf{if }}
\newcommand{\INPUT}{\textbf{Input: }}
\newcommand{\OUTPUT}{\textbf{Output: }}
\newcommand{\RETURN}{\textbf{return }}
\newcommand{\THEN}{\textbf{then }}
\newcommand{\WHILE}{\textbf{while }}

\newlength{\ai}
\newcommand{\ei}{\hspace*{\ai}}

\begin{document}
\title[Computing Graph Roots Without Short Cycles]{Computing Graph Roots Without Short Cycles}

\author[ref1]{B. Farzad}{Babak Farzad}
\address[ref1]{Department of Mathematics, Brock University, Canada.}
\email{bfarzad@brocku.ca}

\author[ref2]{L.C. Lau}{Lap Chi Lau}
\address[ref2]{Department of Computer Science and Engineering, The Chinese University of Hong Kong.}
\email{chi@cse.cuhk.edu.hk}

\author[ref3]{V.B. Le}{Van Bang Le}
\address[ref3]{Universit\"at Rostock, Institut f\"ur Informatik, Germany.}
\email{{le,nn024}@informatik.uni-rostock.de}

\author[ref3,ref4]{N.N. Tuy}{Nguyen Ngoc Tuy$^\dag$}
\address[ref4]{Hong Duc University, Vietnam.}
\email{nntuy@yahoo.com}
\thanks{\dag ~This author is supported by the Ministry of Education and Training, Vietnam, Grant No. 3766/QD-BGD~\&~DT}

% \subjclass{PREFERRED list of ACM classifications}
\thanks{\color{white}{.}}

\begin{abstract}
\noindent
Graph $G$ is the square of graph $H$ if two vertices $x,y$ have an edge in
  $G$ if and only if $x,y$ are of distance at most two in $H$.
Given $H$ it is easy to compute its square $H^2$,
  however Motwani and Sudan proved that it is NP-complete
  to determine if a given graph $G$ is the square of some graph $H$ (of girth $3$).
In this paper we consider the characterization and recognition
  problems of graphs that are squares of graphs of small girth,
  i.e. to determine if $G=H^2$ for some graph $H$ of small girth.
The main results are the following.
\begin{itemize}
\item There is a graph theoretical characterization for graphs that are squares of
  some graph of girth at least $7$.
A corollary is that if a graph $G$ has a square root $H$ of girth at least $7$ then $H$ is unique up to isomorphism.
\item There is a polynomial time algorithm to recognize if $G=H^2$ for some graph $H$ of girth at least $6$.
\item It is NP-complete to recognize if $G=H^2$ for some graph $H$ of girth $4$.
\end{itemize}
These results almost provide a dichotomy theorem for the complexity
  of the recognition problem in terms of girth of the square roots.
The algorithmic and graph theoretical results generalize previous
  results on tree square roots, and provide polynomial time algorithms
  to compute a graph square root of small girth if it exists.
Some open questions and conjectures will also be discussed.
\end{abstract}

\maketitle

\section{Introduction}
%        ============

{\em Root} and {\em root finding} are concepts familiar to most
branches of mathematics.
In graph theory, $H$ is a {\em square root} of $G$
and $G$ is the {\em square} of $H$ if
two vertices $x,y$ have an edge in $G$ if and only if
$x,y$ are of distance at most two in $H$.
Graph square is a basic operation
with a number of results about its properties in the literature. %references
In this paper we are interested in the characterization
and recognition problems of graph squares.
Ross and Harary \cite{RosHar1960} characterized squares of trees and showed
that tree square roots, when they exist, are unique up to isomorphism.
Mukhopadhyay \cite{Muk1967} provided a characterization of graphs which have a
square root, but this is not a good characterization in the sense
that it does not give a short certificate
when a graph does not have a square root.
In fact, such a good characterization may not exist
as Motwani and Sudan proved that it is NP-complete
to determine if a given graph has a square root \cite{MotSud1994}.
%Later Lau and Corneil showed that it remains NP-complete
%to determine if a given graph has a chordal graph square root \cite{LauCor2004}.
On the other hand, there are polynomial time algorithms
to compute the tree square root
\cite{LinSki1995,KeaCor1998,Lau2006,BraLeSri2006,ChaKoLu2006},
a bipartite graph square root \cite{Lau2006},
and a proper interval graph square root \cite{LauCor2004}.

The algorithms for computing tree square roots
and bipartite graph square roots are based on the fact that
the square roots have no cycles and no odd cycles respectively.
Since computing the graph square uses only local information
from the first and the second neighborhood,
it is plausible that there are polynomial time algorithms
to compute square roots that have no short cycles
(locally tree-like), and more generally to compute square roots that
have no short odd cycles (locally bipartite).
The {\em girth} of a graph is the length of a shortest cycle.
In this paper we consider the characterization and recognition
problems of graphs that are squares of graphs of small girth,
i.e. to determine if $G=H^2$ for some graph $H$ of small girth.

The main results of this paper are the following.
In Section~\ref{sec:girthseven} we will provide a good characterization
for graphs that are squares of some graph of girth at least $7$.
This characterization not only leads to a simple algorithm to compute
a square root of girth at least $7$ but also shows such a square root, if it exists, is unique
up to isomorphism.
Then, in Section~\ref{sec:girthsix}, we will present a polynomial time
algorithm to compute a square root of girth at least $6$, or report that none exists.
In Section~\ref{sec:girth3&4} we will show that it is NP-complete to determine if a graph $G$ has
a square root of girth $4$.
Finally, we discuss some open questions and conjectures.

These results almost provide a dichotomy theorem for the complexity
of the recognition problem in terms of girth of the square roots.
The algorithmic and graph theoretical results considerably
generalize previous results on tree square roots.
We believe that our algorithms can be extended to
compute square roots with no short odd cycles (locally bipartite),
and in fact one part of the algorithm for computing square
roots of girth at least $6$ uses only the assumption that
the square roots have no $3$ cycles or $5$ cycles.
Coloring properties of squares in terms of girth of the roots
have been considered in the literature \cite{AloMoh,CraKim,Havet};
our algorithms would allow those results to apply even
though a square root was not known apriori.

{\bf Definitions and notation:}
All graphs considered are finite, undirected and simple. Let $G=(V_G, E_G)$ be a graph.
We often write $xy\in E_G$ for $\{x,y\}\in E_G$. Following \cite{MotSud1994,LauCor2004}, we sometimes
also write $x\leftrightarrow y$ for the adjacency of $x$ and $y$ in the graph in question; this
is particularly the case when we describe reductions in NP-completeness proofs.

The \emph{neighborhood} $N_G(v)$ in $G$
of a vertex $v$ is the set all vertices in $G$ adjacent to $v$ and the
\emph{closed neighborhood} of $v$ in $G$ is $N_G[v] = N_G(v)\cup\{v\}$. Set $\deg_G(v)=|N_G(v)|$,
the \emph{degree} of $v$ in $G$. We call vertices of degree one in $G$ \emph{end-vertices}
of $G$. A \emph{center vertex} of $G$ is one that is adjacent to all other vertices.

Let $d_G(x,y)$ be the length, i.e., number of edges, of a shortest path in $G$ between
$x$ and $y$. Let $G^k=(V_G,E^k)$ with $xy \in E^k$ if and only if $1\le d_G(x,y) \le k$ denote
the {\em $k$-th power of $G$}. If $G=H^k$ then $G$ is the $k$-th power of the graph $H$ and
$H$ is a \emph{$k$-th root} of $G$. Since the power of a graph $H$ is the union of the powers
of the connected components of $H$, we may assume that all graphs considered are connected.

A set of vertices $Q\subseteq V_G$ is called a \emph{clique} in $G$ if every two
distinct vertices in $Q$ are adjacent; a \emph{maximal clique} is a clique that is not
properly contained in another clique. A \emph{stable set} is a set of pairwise non-adjacent
vertices. Given a set of vertices $X\subseteq V_G$, the subgraph induced by $X$ is written
$G[X]$ and $G-X$ stands for $G[V\setminus X]$. If $X=\{a, b, c, \ldots\}$, we write
$G[a, b, c, \ldots]$ for $G[X]$. Also, we often identify a subset of vertices with the
subgraph induced by that subset, and vice versa.

The \emph{girth} of $G$, $\text{girth}(G)$, is the smallest length of a cycle in $G$; in
case $G$ has no cycles, we set $\text{girth}(G)=\infty$. In other words, $G$ has girth
$k$ if and only if $G$ contains a cycle of length $k$ but does not contain any (induced)
cycle of length $\ell=3, \ldots, k-1$.
Note that the girth of a graph can be computed
in $O(nm)$ time, where $n$ and $m$ are the number of vertices, respectively, edges of the input
graph~\cite{ItaRod}.

A complete graph is one in which every two distinct vertices are adjacent; a complete graph on $k$
  vertices is also denoted by $K_k$.
A \emph{star} is a graph with \emph{at least two} vertices that has a center vertex and
  the other vertices are pairwise non-adjacent.
Note that a star contains at least one edge and at least one center vertex;
  the center vertex is unique whenever the star has more than two vertices.

\section{Squares of graphs with girth at least seven}\label{sec:girthseven}
%        =================================================================
In this section, we give a good characterization of graphs that are squares
of a graph of girth at least seven.
Our characterization leads to a simple polynomial-time
recognition for such graphs.
\begin{prop}\label{prop:maxclique}
Let $G$ be a connected, non-complete graph such that $G=H^2$ for some graph $H$. \\
(i)  If girth$(H)\ge 6$ and $v$ is a vertex with $\deg_H(v)\ge 2$ then $N_H[v]$ is a maximal clique in~$G$;\\
(ii) If girth$(H)\ge 7$ and $Q$ is a maximal clique in $G$ then $Q=N_H[v]$ for some vertex $v$ where $\deg_H(v)\ge 2$.
\end{prop}

\proof
(i) Let $v$ be a vertex with $\deg_H(v)\ge 2$. Clearly, $Q=N_H[v]$ is a clique in $G$.
Consider an arbitrary vertex $w$ outside $Q$; in particular, $w$ is non-adjacent in $H$ to $v$.
If $w$ is non-adjacent in $H$ to all vertices in $Q$, then $d_H(w, v) > 2$. If $w$ is adjacent
in $H$ to a vertex $x\in Q-v$, let $y\in Q\setminus\{v, x\}$. Then $N_H[w]\cap N_H[y]=\emptyset$
(otherwise $H$ would contain a cycle of length at most five), hence $d_H(w, y) > 2$. Thus, in
any case, $w$ cannot be adjacent, in $G$, to all vertices in $Q$, and so $Q$ is a maximal
clique in $G$.

(ii) Let $Q$ be a maximal clique in $G$ and $v\in Q$ be a vertex that maximizes $|Q\cap N_H[v]|$.
We prove that $Q=N_H[v]$.
% First, we show that $\deg_H(v)\ge 2$.
% Assume otherwise and let $u$ be the (only) neighbor of $v$ in $H$.
% Then, for every vertex $x$, if $d_H(x, v)\le 2$, then $d_H(x, u)\le 1$.
% Hence $Q\subseteq N_H[u]$, and by the maximality of $Q$, $Q=N_H[u]$.
% Now, since $G$ is not complete, $\deg_H(u)\ge 2$, and so, $|Q\cap N_H[u]|=|N_H[u]|\ge 3 > |Q\cap N_H[v]| = 2$, contradicting the choice of $v$.
It can be seen that by the maximality of $Q$, $\deg_H(v)\ge 2$.
Now, we show that if $w\in Q\setminus N_H[v]$ and $x\in Q\cap N_H[v]$, then $wx\not\in E_H$:
As $w\not\in N_H[v]$, this is clear in case $x=v$. So, let $x\not= v$ and
assume to the contrary that $wx\in E_H$. Then, by the choice of $v$, there exists a vertex
$w'\in Q\setminus N_H[x]$, $w'\in N_H[v]$. Note that $w'x, w'w\not\in E_H$ because $H$ has no
$C_3, C_4$. As $ww'\in E_G\setminus E_H$, there exists a vertex $u\not\in\{w, w', x, v\}$ with
$uw, uw'\in E_H$. But then $H[w, w', x, v, u]$ contains a $C_4$ or $C_5$. Contradiction.

Finally, we show that $Q\subseteq N_H[v]$, and so, by the maximality of $Q$, $Q=N_H[v]$:
Assume otherwise and let $w\in Q\setminus N_H[v]$.
As $wv\in E_G\setminus E_H$, there exists a vertex $x$ such that $xw, xv\in E_H$, and so,
$x\in N_H[v]\setminus Q$. By the maximality of $Q$, $x$ must be non-adjacent (in $G$) to a vertex
$w'\in Q$. In fact, $w'\in Q\setminus N_H[v]$ as $x$ is adjacent in $G$ to every vertex in $N_H[v]$.
Since $w'v\in E_G\setminus E_H$, there exists a vertex $a$ such that
$aw', av\in E_H$; note that $a\not\in\{x,w\}$. Now, if $ww'\in E_H$ then $H[w, w', a, v, x]$
contains a cycle of length at most five. If $ww'\not\in E_H$, let $b$ be a vertex such that
$bw, bw'\in E_H$; possibly $b=a$. Then $H[w, w', a, b, v, x]$ contains a cycle of length at most
six. In any case we have a contradiction, hence $Q\setminus N_H[v]=\emptyset$.\qed

The $5$-cycle $C_5$ and the $6$-cycle $C_6$ show that (i), respectively, (ii) in
Proposition~\ref{prop:maxclique} is best possible with respect to the girth condition of
the root. More generally, the maximal cliques in the square of the subdivision of any complete graph
on $n\ge 3$ vertices do not satisfy Condition (ii).

\begin{definition}\label{defi:forced}
Let $G$ be an arbitrary graph. An edge of $G$ is called \emph{forced} if it is contained in (at least)
two distinct maximal cliques in $G$.
\end{definition}
\begin{proposition}\label{prop:forced}
Let $G$ be a connected, non-complete graph such that $G=H^2$ for some graph $H$ with girth at least
seven, and let $F$ be the subgraph of $G$ consisting of all forced edges of $G$. Then\\
(i) $F$ is obtained from $H$ by deleting all end-vertices in $H$;\\
(ii) for every maximal clique $Q$ in $G$, $F[Q\cap V_F]$ is a star; and\\
(iii) every vertex in $V_G-V_F$ belongs to exactly one maximal clique in $G$.
\end{proposition}
\proof First we observe that $xy$ is a forced edge in $G$ iff $xy$ is an edge in $H$
  with $\deg_H(x)\ge 2$ and $\deg_H(y)\ge 2$.
% We first make the following two observations:
%
% 1) Consider a forced edge $xy$ in $G$. Let $Q_1\not= Q_2$ be two maximal cliques in $G$ containing
% $xy$. By Proposition~\ref{prop:maxclique}, there exist vertices $v_i$, $i=1,2$, with $\deg_H(v_i)\ge 2$
% and $Q_i=N_H[v_i]$. As $Q_1\not= Q_2$, $v_1\not=v_2$. As $x,y\in N_H[v_1]\cap N_H[v_2]$ and $H$ has no
% $C_3, C_4$, $\{x,y\}=\{v_1, v_2\}$ and $xy=v_1v_2\in E_H$. Thus, every forced edge $xy$ in $G$ is an
% edge in $H$ with $\deg_H(x)\ge 2$ and $\deg_H(y)\ge 2$.
%
% 2) Let $xy$ be an edge in $H$. If $x$ or $y$ is an end-vertex in $H$, then clearly $xy$ belongs to
% exactly one maximal clique in $G$, hence $xy$ is not a forced edge in $G$. If $\deg_H(x)\ge 2$ and
% $\deg_H(y)\ge 2$, then by Proposition~\ref{prop:maxclique}, $N_H[x]$ and $N_H[y]$ are two (distinct)
% maximal cliques in $G$ containing $xy$, hence $xy$ is a forced edge in $G$.
Now, (i) follows directly from the above observations. For (ii),
consider a maximal clique $Q$ in $G$. By Proposition~\ref{prop:maxclique}, $Q=N_H[v]$ for
some vertex $v$ with $\deg_H(v)\ge 2$. Let $X$ be the set of all neighbors of $v$ in $H$ that are
end-vertices in $H$ and $Y=N_H(v)\setminus X$. Since $G$ is not complete, $Y\not=\emptyset$. By (i),
$X\cap V_F=\emptyset$, hence $F[Q\cap V_F]=F[\{v\}\cup Y]$ which implies (ii).
For (iii), consider a vertex $u\in V_G-V_F$ and a maximal clique $Q$ containing $u$.
Then, $u$ cannot belong to $Y$ and therefore $Q$ is the only maximal clique containing $u$. \qed

We now are able to characterize squares of graphs with girth at least seven as follows.
\begin{theorem}\label{thm:girthseven}
Let $G$ be a connected, non-complete graph. Let $F$ be the subgraph of $G$ consisting of all forced
edges in $G$. Then $G$ is the square of a graph with girth at least seven if and only if the following
conditions hold.\\
(i)   Every vertex in $V_G-V_F$ belongs to exactly one maximal clique in $G$. \\
(ii)  Every edge in $F$ belongs to exactly two distinct maximal cliques in $G$.\\
(iii) Every two non-disjoint edges in $F$ belong to a common maximal clique in $G$.\\
(iv)  For each maximal clique $Q$ of $G$, $F[Q\cap V_F]$ is a star.\\
(v)   $F$ is connected and has girth at least seven.
\end{theorem}
\proof For the only if-part, (ii) and (iii) follow easily from Proposition~\ref{prop:maxclique},
and (i), (iv) and (v) follow directly from Proposition~\ref{prop:forced}.

For the if-part, let $G$ be a connected graph satisfying (i) -- (v). We will construct a spanning
subgraph $H$ of $G$ with girth at least seven such that $G=H^2$ as follows. For each edge $xy$ in
$F$ let, by (ii) and (iv), $Q\not= Q'$ be \emph{the} two maximal cliques in $G$ with $Q\cap Q'=\{x,y\}$.
Let, without loss of generality, $|Q\cap V_F|\ge |Q'\cap V_F|$.
% By (iv),
Assuming $x$ is a center vertex of the star $F[Q\cap V_F]$, then $y$ is a center vertex of the star $F[Q'\cap V_F]$:
Otherwise, by (iv), $x$ is the center vertex of the star $F[Q'\cap V_F]$ and
  there exists some $y'\in Q'\cap V_F$ such that $yy'\not\in F$; note that $xy'\in F$ (by (iv)).
As $|Q\cap V_F|\ge |Q'\cap V_F|$, there is an edge $xz\in F-xy$ in $Q-Q'$. By (iii), $zy'\in E_G$.
Now, as $Q'$ is maximal, the maximal clique $Q''$ containing $x, y, z, y'$ is different from $Q'$.
But then $\{y,y'\}\subseteq Q'\cap Q''$, i.e., $yy'\in F$, hence $F$ contains a triangle $xyy'$,
contradicting (v).

Thus, assuming $x$ is a center vertex of the star $F[Q\cap V_F]$, $y$ is a center vertex of the
star $F[Q'\cap V_F]$. Then put the edges $xq$, $q\in Q-x$, and $yq'$, $q'\in Q'-y$, into $H$.

By construction, $F\subseteq H\subseteq G$ and by (i),
\begin{equation}\label{eqn:2}
\text{for all vertices $u\in V_H\setminus V_F,~\deg_H(u)=1$},
\end{equation}
\begin{equation}\label{eqn:3}
\forall v\in V_F, \forall a, b \in V_H\text{ with }va, vb\in E_H: \text{ $a$ and $b$ belong to
the same clique in $G$}.
\end{equation}
Furthermore, as every maximal clique in $G$ contains a forced edge (by (iv)), $H$ is a spanning
subgraph of $G$. Moreover, $F$ is an induced subgraph of $H$: Consider an edge $xy\in E_H$ with
$x, y\in V_F$. By construction of $H$, $x$ or $y$ is a center vertex of the star $F[Q\cap V_F]$
for some maximal clique $Q$ in $G$. Since $x, y\in V_F$, $xy$ must be an edge of this star, i.e.,
$xy\in E_F$. Thus, $F$ is an induced subgraph of $H$. In particular, by (\ref{eqn:2}) and (v),
$H$ is connected and $\text{girth}(H)=\text{girth}(F)\ge 7$.

Now, we complete the proof of Theorem~\ref{thm:girthseven} by showing that $G=H^2$.
Let $uv\in E_G\setminus E_H$ and let $Q$ be a maximal clique in $G$ containing
$uv$. By (iv), $Q$ contains a forced edge $xy$ and $x$ or $y$ is a center vertex of the star
$F[Q\cap V_F]$. By construction of $H$, $xu$ and $xv$, or else $yu$ and $yv$ are edges of $H$,
hence $uv\in E_{H^2}$. This proves $E_G\subseteq E_{H^2}$.
Now, let $ab\in E_{H^2}\setminus E_H$. Then there exists a vertex $x$ such that
$xa, xb\in E_H$. By (\ref{eqn:2}), $x\in V_F$, and by (\ref{eqn:3}), $ab\in E_G$.
This proves $E_{H^2}\subseteq E_G$. \qed

\begin{corollary}\label{coro:girthseven}
Given a graph $G=(V_G, E_G)$, it can be recognized in $O(|V_G|^2\cdot |E_G|)$ time if $G$ is the
square of a graph $H$ with girth at least seven. Moreover, such a square root, if any, can be
computed in the same time.
\end{corollary}
\proof Note that by Proposition~\ref{prop:maxclique}, any square of an $n$-vertex graph with girth
at least seven has at most $n$ maximal cliques. Now, to avoid triviality, assume $G$ is connected
and non-complete. We first use the algorithm in \cite{TIAS} to list the maximal cliques in $G$ in
time $O(n^2m)$. If there are more than $n$ maximal cliques, $G$ is not the square of any graph
with girth at least seven. Otherwise, compute the forced edges of $G$ to form the subgraph $F$ of
$G$. This can be done in time $O(n^2)$ in an obvious way.
Conditions (i) -- (v) in Theorem~\ref{thm:girthseven} then can
be tested within the same time bound, as well as the square root $H$, in case all conditions are
satisfied, according to the proof of Theorem~\ref{thm:girthseven}. \qed

\begin{corollary}\label{coro:unique}
The square roots with girth at least seven of squares of graphs with girth at least seven are
unique, up to isomorphism.
\end{corollary}
\proof Let $G$ be the square of some graph $H$ with girth $\ge 7$. If $G$ is complete, clearly,
every square root with girth $\ge 6$ of $G$ must be isomorphic to the star $K_{1, n-1}$ where
$n$ is the vertex number of $G$.

Thus, let $G$ be non-complete, and let $F$ be the subgraph of $G$ formed by the forced edges. If
$F$ has only one edge, $G$ clearly consists of exactly two maximal cliques, $Q_1$, $Q_2$, say,
and $Q_1\cap Q_2$ is the only forced edge of $G$. Then, it is easily seen that every square root
with girth $\ge 6$ of $G$ must be isomorphic to the double star $T$ having center edge $v_1v_2$
and $\deg_T(v_i)=|Q_i|$.

So, assume $F$ has at least two edges. Then for each two maximal cliques
$Q, Q'$ in $G$ with $Q\cap Q'=\{x,y\}$, $x$ or $y$ is the unique center vertex of the star
$F[V_F\cap Q]$ or $F[V_F\cap Q']$. Hence, for any end-vertex $u$ of $H$, i.e., $u\in V_G-V_F$,
the neighbor of $u$ in $F$ is unique. Since $F$ is the graph resulting from $H$ by deleting
all end-vertices, $H$ is therefore unique. \qed

\subsection{Further Considerations}\label{sec:girth-k}

Squares of bipartite graphs can be recognized in $O(\Delta\cdot M(n))$ time in \cite{Lau2006},
where $\Delta=\Delta(G)$ is the maximum degree of the $n$-vertex input graph $G$ and $M(n)$ is
the time needed to perform the multiplication of two $n\times n$-matrices. However, no good
characterization is known so far. As bipartite graphs with girth at least seven are exactly the
$(C_4,C_6)$-free bipartite graphs, we immediately have:

\begin{corollary}\label{coro:C4C6freebip}
Let $G$ be a connected, non-complete graph. Let $F$ be the subgraph of $G$ consisting of all forced
edges in $G$. Then $G$ is the square of a $(C_4,C_6)$-free bipartite graph if and only if the following
conditions hold.\\
(i)  Every vertex in $V_G-V_F$ belongs to exactly one maximal clique in $G$.\\
(ii)  Every edge in $F$ belongs to exactly two distinct maximal cliques in $G$.\\
(iii) Every two non-disjoint edges in $F$ belong to the same maximal clique in $G$.\\
(iv)  For each maximal clique $Q$ of $G$, $F[Q\cap V_F]$ is a star.\\
(v)   $F$ is a connected $(C_4,C_6)$-free bipartite graph.\\
Moreover, squares of $(C_4,C_6)$-free bipartite graphs can be recognized in $O(n^2m)$ time, and the
$(C_4, C_6)$-free square bipartite roots of such squares are unique, up to isomorphism.
\end{corollary}

Using the results in this section, we obtain a new characterization
for tree squares that allow us to derive the known results on tree
square roots easily.
% See Appendix~\ref{sec:trees} for details.

It was shown in \cite{LinSki1995} that \textsc{clique} and \textsc{stable set} remain NP-complete on
squares of graphs (of girth three). Another consequence of our results is.

\begin{corollary}\label{coro:CLIQUE}
The weighted version of \textsc{clique} can be solved in $O(n^2m)$ time on squares of graphs
with girth at least 7, where $n$ and $m$ are the number of vertices, respectively, edges of the
input graph.
\end{corollary}
\proof Let $G=(V_G, E_G)$ be the square of some graph with girth at least seven.
By Proposition~\ref{prop:maxclique}, $G$ has $O(|V_G|)$ maximal cliques. By \cite{TIAS}, all maximal
cliques in $G$ then can be listed in time $O(|V_G|\cdot |E_G|\cdot |V_G|)$.\qed

In \cite{HorKil}, it was shown that \textsc{stable set} is even NP-complete on squares of the subdivision
of some graph (i.e. the squares of the total graph of some graph). As the subdivision of a graph has girth
at least six, \textsc{stable set} therefore is NP-complete on squares of graphs with girth at least six.

\section{Squares of graphs with girth at least six}\label{sec:girthsix}
%=====================================================================
In this section we will show that squares of graphs with girth at least six can be recognized
efficiently. Formally, we will show that the following problem

\textsc{square of graph with girth at least six}\\[1ex]
\begin{tabular}{ll}
{\em Instance:}& A graph $G$.\\
{\em Question:}& Does there exist a graph $H$ with girth at least $6$ such that $G=H^2$?\\
\end{tabular}

is polynomially solvable (Theorem~\ref{thm:girth6}).

Similar to the algorithm in \cite{Lau2006},
our recognition algorithm consists of two steps.
The first step (subsection~\ref{c3c5-free}) is
to show that if we fix a vertex $v\in V$ and a subset $U \subseteq N_G(v)$, then there is at
most one $\{C_3,C_5\}$-free (locally bipartite)
square root graph $H$ of $G$ with $N_H(v) = U$.
Then, in the second step (subsection~\ref{fix-edge}), we show that if we fix an edge $e = uv\in E_G$,
then there are at most two possibilities of $N_H(v)$ for a square root $H$ with girth at least $6$.
Furthermore, both steps can be implemented efficiently,
and thus it will imply
that \textsc{square of graph with girth at least six} is polynomially solvable.

\subsection{Square root with a specified neighborhood}\label{c3c5-free}
This subsection deals with the first auxiliary problem.

\textsc{$\{C_3,C_5\}$-free square root with a specified neighborhood}\\[1ex]
\begin{tabular}{l l}
{\em Instance:}& A graph $G$, $v \in V_G$ and $U \subseteq N_G(v)$. \\
{\em Question:}& Does there exist a $\{C_3,C_5\}$-free graph $H$ such that $H^2=G$ and $N_H(v) = U$?\\
\end{tabular}

An efficient recognition algorithm for \textsc{$\{C_3,C_5\}$-free square root with a specified neighborhood}
relies on the following fact.

\begin{lemma}\label{lem:c3c5-free}
Let $G=H^2$ for some $\{C_3,C_5\}$-free graph $H$. Then, for all vertices $x\in V$ and
all vertices $y\in N_H(x)$, $N_H(y)=N_G(y)\cap\big(N_G[x]\setminus N_H(x)\big)$.
\end{lemma}
\proof First, consider an arbitrary vertex $w\in N_H(y)-x$. Clearly, $w\in N_G(y)$, as well
 $w\in N_G(x)$. Also, since $H$ is $C_3$-free, $wx\not\in E_H$. Thus
 $w\in N_G(y)\cap\big(N_G(x)\setminus N_H(x)\big)$.

 Conversely, let $w$ be an arbitrary vertex in $N_G(y)\cap\big(N_G[x]\setminus N_H(x)\big)$.
 Assuming $wy\not\in E_H$, then $w\not= x$ and there exist vertices $z$ and $z'$ such that
 $zx, zw\in E_H$ and $z'y, z'w\in E_H$. As $H$ is $C_3$-free, $zy\not\in E_H$, $z'x\not\in E_H$,
 and $zz'\not\in E_H$. But then $x, y, w, z$ and $z'$ induce a $C_5$ in $H$, a contradiction.
 Thus $w\in N_H(y)$.\qed

Recall that $M(n)$ stands for the time needed to perform a matrix multiplication of two $n\times n$ matrices;
currently, $M(n) =O(n^{2.376})$.

\begin{theorem}\label{thm:c3c5-free}
\textsc{$\{C_3,C_5\}$-free square root with a specified neighborhood} has at most one solution.
The unique solution, if any, can be constructed in time $O(M(n))$.
\end{theorem}
\proof Given $G$, $v\in V_G$ and $U\subseteq N_G(v)$, assume $H$ is a $\{C_3,C_5\}$-free
square root of $G$ such that $N_H(v)=U$. Then, by Lemma~\ref{lem:c3c5-free},
the neighborhood in $H$ of each vertex $u\in U$ is uniquely determined by
$N_H(u)=N_G(u)\cap\big(N_G[v]\setminus U\big)$. By repeatedly applying Lemma~\ref{lem:c3c5-free}
for each $v'\in U$ and $U'=N_H(v')$ and noting that all considered graphs are
connected, we can conclude that $H$ is unique.

Lemma~\ref{lem:c3c5-free} also suggests the following BFS-like procedure, Algorithm~1
below, for constructing the $\{C_3,C_5\}$-free square root $H$ of $G$ with $U=N_H(v)$,
if any.

It can be seen, by construction, that $H$ is $\{C_3,C_5\}$-free, and thus the
correctness of Algorithm~1 follows from Lemma~\ref{lem:c3c5-free}. Moreover,
since every vertex is enqueued at most once, lines 1--13 take $O(m)$ steps,
$m=|E_G|$. Checking if $G=H^2$ (line~14) takes $O(M(n))$ steps, $n=|V_G|$.\qed

\begin{center}
    ALGORITHM 1 \\[1ex]
    \fbox{\quad
       \begin{minipage}{\textwidth}
        \begin{tabbing}
          \OUTPUT\=\kill
          \INPUT \> A graph $G$, a vertex $v \in V_G$ and a subset $U \subseteq N_G(v)$. \\
          \OUTPUT\> A $\{C_3,C_5\}$-free graph $H$ with $H^2=G$ and $N_H(v) = U$,\\
                 \> or else `NO' if such a square root $H$ of $G$ does not exist.\\\\

          (Nr.)\=\ei\=\ei\=\ei\=\ei\=\ei\=\ei\=\ei\=\ei\=\ei\=\ei\=\ei\=\ei\=\kill

           1.\> Add all edges $vu$, $u\in U$, to $E_H$\\
           2.\> $Q \leftarrow \emptyset$\\
           3.\> \FOR each $u\in U$ \DO\\
           4.\>\>\>\>  $\mathrm{enqueue}(Q,u)$\\
           5.\>\>\>\>  $\mathrm{parent}(u) \leftarrow v$\\
           6.\> \WHILE $Q\not=\emptyset$ \DO\\
           7.\>\>\>\> $u \leftarrow\mathrm{dequeue}(Q)$\\
           8.\>\>\>\> set $W:= N_G(u)\cap\big(N_G(\mathrm{parent}(u))\setminus N_H(\mathrm{parent}(u))\big)$\\
           9.\>\>\>\> \FOR each $w\in W$ \DO\\
          10.\>\>\>\>\>\>\> add $uw$ to $E_H$\\
          11.\>\>\>\>\>\>\> \IF $\mathrm{parent}(w) = \emptyset$\\
          12.\>\>\>\>\>\>\> \THEN $\mathrm{parent}(w) \leftarrow u$\\
          13.\>\>\>\>\>\>\>\>\>\>\> $\mathrm{enqueue}(Q,w)$\\
          14.\> \IF $G=H^2$ \THEN \RETURN $H$\\
          15.\>\>\> \ELSE \RETURN `NO'
        \end{tabbing}
      \end{minipage}
    \quad}
  \end{center}

\subsection{Square root with a specified edge}\label{fix-edge}
This subsection discusses the second auxiliary problem.

\textsc{girth $\ge 6$ root graph with one specified edge}\\[1ex]
\begin{tabular}{l l}
{\em Instance:}& A graph $G$ and an edge $xy \in E_G$.\\
{\em Question:}& Does there exist a graph $H$ with girth at least six such that $H^2 = G$\\
               & and $xy\in E_H$?\\
\end{tabular}

The question is easy if $|G|\leq 2$.
So, for the rest of this section, assume that $|G| > 2$.
Then, we will reduce this problem to
\textsc{$\{C_3,C_5\}$-free square root with a specified neighborhood}.
Given a graph $G$ and an edge $xy$ of $G$, write $C_{xy} = N_G(x) \cap N_G(y)$,
i.e., $C_{xy}$ is the set of common neighbors of $x$ and $y$ in $G$.

\begin{lemma}\label{lem:fix-edge1}
Suppose $H$ is of girth at least $6$, $xy \in E_H$ and $H^2 = G$. Then
$G[C_{xy}]$ has at most two connected components. % and each of them is a clique.
Moreover, if $A$ and $B$ are the connected components of $G[C_{xy}]$
$($one of them maybe empty$)$ then (i) $A=N_H(x)-y$ and $B=N_H(y)-x$, or (ii) $B=N_H(x)-y$ and $A=N_H(y)-x$.
\end{lemma}

% \begin{proof} Set $X=N_H(x)-y$ and $Y=N_H(y)-x$. Notice that $X$ or $Y$ (but not both) may be empty.
% First we show that $X \cup Y = C_{xy}$.
% Consider an arbitrary vertex $v \in C_{xy}$;
% we claim that $v$ is either in $X$ or $Y$.
% Otherwise, there is a length 2 path from $v$ to $x$
% and a length 2 path from $v$ to $y$,
% which implies that there is either a 3-cycle or a 5-cycle, a contradiction.
% So we have $C_{xy} \subseteq X \cup Y$.
%
% On the other hand, consider an arbitrary vertex $u \in X$.
% It is obvious that $u \in N_{H^2}(x)$.
% Also, since $xy \in E_H$, $u \in N_{H^2}(y)$.
% A similar argument applies if $u \in Y$.
% Therefore, $u \in N_{H^2}(x) \cap N_{H^2}(y)$.
% Since $H^2=G$, $u \in C_{xy}$.
% Hence $X \cup Y = C_{xy}$.
%
% Next, observe that $X$ and $Y$ induce cliques in $H^2$
% and thus in $G$. Moreover, $X\cap Y=\emptyset$ (as $H$ has no $3$-cycle)
% and no vertex in $X$ is adjacent in $H$ to a vertex in $Y$ (as $H$ has no $4$-cycle).
% Now, no vertex $u\in X$ is adjacent in $G$ to a vertex $w\in Y$: Otherwise, there is
% a vertex $v\notin X\cup Y$ adjacent in $H$ to $u$ and to $w$, implying that
% $x, y, u, w, v$ induce a $5$-cycle in $H$, a contradiction.
%
% Thus, the cliques $G[X]$ and $G[Y]$ are exactly the connected components of $G[C_{xy}]$
% and the lemma follows.
% \end{proof}

By Lemma~\ref{lem:fix-edge1}, we can solve \textsc{girth $\ge 6$ root graph with one specified edge}
as follows: Compute $C_{xy}$. If $G[C_{xy}]$ has more than two connected components, there is no solution.
If $G[C_{xy}]$ is connected, solve
\textsc{$\{C_3,C_5\}$-free square root with a specified neighborhood} for inputs $I_1=(G, v=x,
U=C_{xy}+y)$ and $I_2=(G, v=y, U=C_{xy}+x)$. If, for $I_1$ or $I_2$, Algorithm~1 outputs $H$
and $H$ is $C_4$-free, then $H$ is a solution. In other cases there is no solution. If $G[C_{xy}]$
has two connected components, $A$ and $B$, solve
\textsc{$\{C_3,C_5\}$-free square root with a specified neighborhood} for inputs $I_1=(G, v=x,
U=A+y)$, $I_2=(G, v=x, U=B+y)$, $I_3=(G, v=y, U=A+x)$, $I_4=(G, v=y, U=B+x)$, and make a decision
similar as before. In this way, checking if a graph is $C_4$-free is the most expensive step,
and we obtain

\begin{theorem}\label{thm:fix-edge}
\textsc{girth $\ge 6$ root graph with one specified edge} can be solved in time $O(n^4)$.
\end{theorem}

Let $\delta=\delta(G)$ denote the minimum vertex degree in $G$. Now we can state the main result of this
section as follows.
\begin{theorem}\label{thm:girth6}
\textsc{square of graph with girth at least six} can be solved in time $O(\delta\cdot n^4)$.
\end{theorem}
\proof Given $G$, let $x$ be a vertex of minimum degree in $G$. For each vertex $y\in N_G(x)$
check if the instance $(G, xy\in E_G)$ for \textsc{girth $\ge 6$ root graph with one specified edge}
has a solution.\qed

\section{Squares of graphs with girth four}\label{sec:girth3&4}
%        ===============================================================
Note that the reductions for proving the NP-completeness results by Motwani and Sudan~\cite{MotSud1994} show that recognizing squares of graphs with
girth three is NP-complete.
In this section we show that the following problem is NP-complete.

\textsc{square of graph with girth four}\\[1ex]
\begin{tabular}{l l}
{\em Instance:}& A graph $G$.\\
{\em Question:}& Does there exist a graph $H$ with girth $4$ such that $G=H^2$?\\
\end{tabular}

Observe that \textsc{square of graph with girth four} is in NP. We will reduce the following
NP-complete problem \textsc{set splitting} \cite[Problem SP4]{GarJoh}, also known as
\textsc{hypergraph 2-colorability}, to it.

\textsc{set splitting}\\[1ex]
\begin{tabular}{l l}
{\em Instance:}& Collection $D$ of subsets of a finite set $S$.\\
{\em Question:}& Is there a partition of $S$ into two disjoint subsets $S_1$ and $S_2$ such that\\
               & each subset in $D$ intersects both $S_1$ and $S_2$?\\
\end{tabular}

Our reduction is a modification of the reductions
for proving the NP-completeness of \textsc{square of chordal graph} \cite[Theorem 3.5]{LauCor2004}
and for \textsc{cube of bipartite graph} \cite[Theorem 7.6]{Lau2006}.
%The details are in Appendix~\ref{sec:NPc}.
We also apply the tail structure of a vertex $v$, first described in~\cite{MotSud1994},
to ensure that $v$ has the same neighbors in any square root $H$ of $G$.

\begin{figure}[H]
  \begin{center}
    \fbox{\quad
    \input{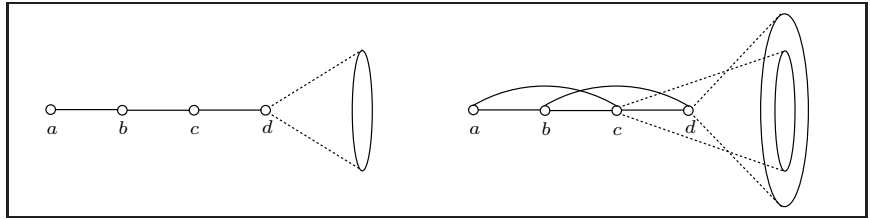}
    \quad}
    \caption{Tail in $H$ (left) and in $G=H^2$ (right)}
    \label{tail}
  \end{center}
\end{figure}

\begin{lemma}[\cite{MotSud1994}]\label{lem:tail}
Let $a, b, c$ be vertices of a graph $G$ such that (i) the only neighbors of $a$ are $b$ and $c$,
(ii)~the only neighbors of $b$ are $a, c$, and $d$, and
(iii)~$c$ and $d$ are adjacent.
Then the neighbors, in $V_G-\{a,b,c\}$, of $d$ in any square root of $G$ are the same as the
neighbors, in $V_G-\{a,b,d\}$, of $c$ in $G$; see Figure~$\ref{tail}$.
\end{lemma}

We now are going to describe the reduction.
Let $S=\{u_1, \ldots, u_n\}$, $D=\{d_1, \ldots, d_m\}$ where $d_j\subseteq S$, $1\le j \le m$,
be an instance of \textsc{set splitting}. We construct an instance $G=G(D, S)$ for
\textsc{square of graph with girth four} as follows.

The vertex set of graph $G$ consists of:\\
\textbf{(I)} $U_i$, $1\le i\le n$. Each `element vertex' $U_i$ corresponds to the element $u_i$ in $S$.\\
\textbf{(II)} $D_j$, $1\le j\le m$. Each `subset vertex' $D_j$ corresponds to the subset $d_j$ in $D$.\\
\textbf{(III)} $D_j^1, D_j^2, D_j^3$, $1\le j\le m$. Each three `tail vertices' $D_j^1, D_j^2, D_j^3$ of
       the subset vertex $D_j$ correspond to the subset $d_j$ in $D$.\\
\textbf{(IV)} $S_1, S_1', S_2, S_2'$, four `partition vertices'.\\
\textbf{(V)} $X$, a `connection vertex'.

The edge set of graph $G$ consists of:\\
\textbf{(I)} Edges of tail vertices of subset vertices:\\
       For all $1\le j\le m$: $D_j^3\leftrightarrow D_j^2$, $D_j^3\leftrightarrow D_j^1$, $D_j^2\leftrightarrow D_j^1$, $D_j^2\leftrightarrow D_j$,
       $D_j^1\leftrightarrow D_j$, and for all $i$, $D_j^1\leftrightarrow U_i$ whenever $u_i\in d_j$.\\
\textbf{(II)} Edges of subset vertices:\\
       For all $1\le j\le m$: $D_j\leftrightarrow S_1$, $D_j\leftrightarrow S_1'$, $D_j\leftrightarrow S_2$, $D_j\leftrightarrow S_2'$,
       $D_j\leftrightarrow X$, $D_j\leftrightarrow U_i$ for all $i$, and $D_j\leftrightarrow D_k$ for all $k$ with $d_j\cap d_k\not=\emptyset$.\\
\textbf{(III)} Edges of element vertices:\\
       For all $1\le i\le n$: $U_i\leftrightarrow X$, $U_i\leftrightarrow S_1$, $U_i\leftrightarrow S_2$, $U_i\leftrightarrow S_1'$, $U_i\leftrightarrow S_2'$,
       and $U_i\leftrightarrow U_{i'}$ for all $i'\not= i$.\\
\textbf{(IV)} Edges of partition vertices:\\
       $S_1\leftrightarrow X$, $S_1\leftrightarrow S_1'$, $S_1\leftrightarrow S_2'$,
       $S_2\leftrightarrow X$, $S_2\leftrightarrow S_1'$, $S_2\leftrightarrow S_2'$,
       $S_1'\leftrightarrow X$, $S_2'\leftrightarrow X$.

\begin{figure}[ht]%[H]
  \begin{center}
    \fbox{\quad
    \input{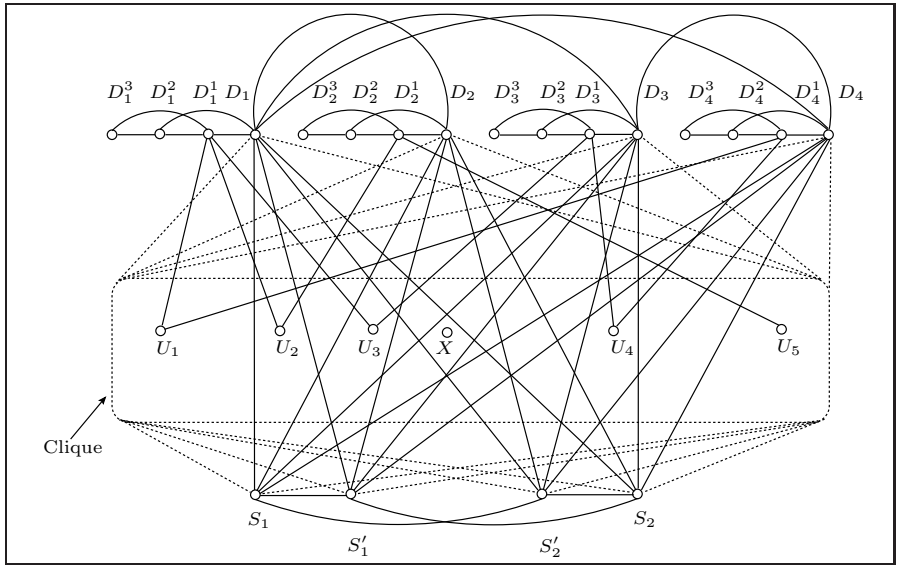}
    \quad}
    \caption{An example of $G$}
    \label{sq-girth4}
  \end{center}
\end{figure}

Clearly, $G$ can be constructed from $D, S$ in polynomial time. For an illustration, given
$S=\{u_1, u_2, u_3, u_4, u_5\}$ and $D=\{d_1, d_2, d_3, d_4\}$ with $d_1=\{u_1,u_2,u_3\}$,
$d_2=\{u_2,u_5\}$, $d_3=\{u_3,u_4\}$, and $d_4=\{u_1,u_4\}$, the graph $G$ is depicted in
Figure~\ref{sq-girth4}. In the figure, the two dotted lines from a vertex to the clique
$\{U_1, U_2, U_3, U_4, U_5, X\}$ mean that the vertex is adjacent to all vertices in that clique.

Note that, apart from the three vertices $X, S_1'$, and $S_2'$ (or, symmetrically, $X, S_1$, and
$S_2$), our construction is the same as those in \cite[\S 3.1.1]{LauCor2004}. While $S_1$ and $S_2$
will represent a partition of the ground set $S$ (Lemma~\ref{lem:girth4-ss}), the vertices $X, S_1'$,
and $S_2'$ allow us to make a square root of $G$ being $C_3$-free (Lemma~\ref{lem:ss-girth4}).

% The proof of Lemmas~\ref{lem:ss-girth4} and \ref{lem:girth4-ss} are in Appendix~\ref{sec:NPc}.

\begin{lemma}\label{lem:ss-girth4}
If there exists a partition of $S$ into two disjoint subsets $S_1$ and $S_2$ such that
each subset in $D$ intersects both $S_1$ and $S_2$, then there exists a graph $H$
with girth four such that $G=H^2$.
\end{lemma}

\begin{figure}[H]
  \begin{center}
    \fbox{\quad
    \input{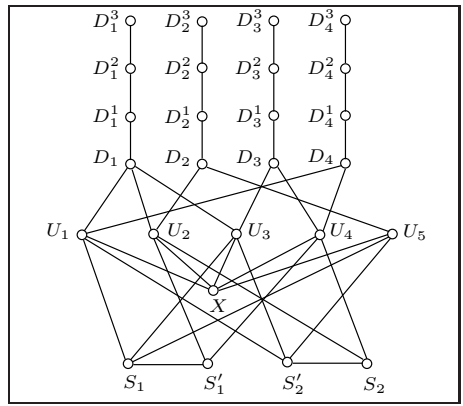}
    \quad}
    \caption{An example of root $H$ with girth $4$}
    \label{root-girth4}
  \end{center}
\end{figure}

In the above example, $S_1=\{u_1,u_3,u_5\}$ and $S_2=\{u_2,u_4\}$ is a possible legal partition
of $S$. The corresponding graph $H$ constructed in the proof of Lemma~\ref{lem:ss-girth4} is depicted
in Figure~\ref{root-girth4}.

\begin{lemma}\label{lem:girth4-ss}
If $H$ is a square root of $G$, then there exists a partition of
$S$ into two disjoint subsets $S_1$ and $S_2$ such that each subset in $D$ intersects
both $S_1$ and $S_2$.
\end{lemma}

Note that in Lemma~\ref{lem:girth4-ss} above we did not require that $H$ has girth four.
Thus, any square root of $G$--particularly, any square root with girth four--will tell us how to
do set splitting. Together with Lemma~\ref{lem:ss-girth4} we conclude:
\begin{theorem}
\textsc{square of graph with girth four} is NP-complete.
\end{theorem}

\section{Conclusion and open problems}
%        ============================
We have shown that squares of graphs with girth at least six can be recognized in polynomial time.
We have found a good characterization for squares of graphs with girth at least seven that
gives a faster recognition algorithm in this case.
For squares of graphs with girth at most four we have shown that recognizing the squares of
such graphs is NP-complete.

The complexity status of computing square root with girth (exactly) five is not yet determined.
However, we believe that this problem should be efficiently solvable.
Also, we believe that the algorithm to compute a square root of girth $6$
can be extended to compute a square root with no $C_3$ or $C_5$.
More generally, let $k$ be a positive integer and consider the following problem.

\textsc{$k$-power of graph with girth $\ge 3k-1$}\\[1ex]
\begin{tabular}{l l}
{\em Instance:}& A graph $G$.\\
{\em Question:}& Does there exist a graph $H$ with girth $\ge 3k-1$ such that $G=H^k$?\\
\end{tabular}

\begin{conjecture}\label{conj}
\textsc{$k$-power of graph with girth $\ge 3k-1$} is polynomially solvable.
\end{conjecture}

The truth of the above conjecture together with the results in this paper would imply a
complete dichotomy theorem: \textsc{squares of graphs of girth $g$} is polynomial if
$g\ge 5$ and NP-complete otherwise.


\begin{thebibliography}{99}
%     ====================
\bibitem{AgnGreHal}
  Geir Agnarsson, Raymond Greenlaw, Magn\'us M. Halld\'orsson,
  On powers of chordal graphs and their colorings,
  {\sl Congressus Numer.} 144 (2000) 41--65.

\bibitem{AloMoh}
  Noga Alon, Bojan Mohar,
  The chromatic number of graph powers,
  {\sl Combinatorics, Probability and Computing} 11 (2002) 1--10.

\bibitem{BraLeSri2006}
  Andreas Brandst\"adt, Van Bang Le, and R. Sritharan,
  Structure and linear time recognition of $4$-leaf powers,
  {\sl ACM Transactions on Algorithms}, to appear.

\bibitem{ChaKoLu2006}
  Maw-Shang Chang, Ming-Tat Ko, and Hsueh-I Lu,
  Linear time algorithms for tree root problems,
  {\sl Lecture Notes in Computer Science}, 4059 (2006) 411--422.

\bibitem{CraKim}
  Daniel W. Cranston, Seog-Jin Kim,
  List-coloring the square of a subcubic graph,
  {\sl J. Graph Theory} 57 (2007) 65--87.

\bibitem{DahDuc1987}
  Elias Dahlhaus, P.~Duchet,
  On strongly chordal graphs,
  {\sl Ars Combin.} 24 B (1987) 23--30.

\bibitem{EscMonRoj1974}
  F. Escalante, L. Montejano, and T. Rojano,
  Characterization of $n$-path graphs and of graphs having $n$th root,
  {\sl J. Combin. Theory B} 16 (1974) 282--289.

\bibitem{GarJoh}
  Michael R. Garey, David S. Johnson,
  {\sl Computers and Intractability--A Guide to the Theory of NP-Completeness},
  Freeman, New York (1979), twenty-third printing 2002.

\bibitem{Golumbic}
  Martin C. Golumbic,
  {\sl Algorithmic Graph Theory and Perfect Graphs},
  Academic Press, New York (1980).

\bibitem{HarKarTut1967}
  Frank Harary, R.M. Karp, and W.T. Tutte,
  A criterion for planarity of the square of a graph,
  {\sl J. Combin. Theory} 2 (1967) 395--405.

\bibitem{Havet}
  Fr\'ed\'eric Havet,
  Choosability of the square of planar subcubic graphs with large girth,
  {\sl Discrete Math.}, to appear.%{\sl INRIA Technical Report} No. 5800, 2006.

\bibitem{HorKil}
  J.D. Horton, K. Kilakos,
  Minimum edge dominating sets,
  {\sl SIAM J. Discrete Math.} 6 (1993) 375--387.

\bibitem{ItaRod}
  Alon Itai, Michael Rodeh,
  Finding a minimum circuit in a graph,
  {\sl SIAM J. Computing} 7 (1978) 413--423.

\bibitem{KeaCor1998}
  Paul E.~Kearney, Derek G.~Corneil,
  Tree powers,
  {\sl J. Algorithms} 29 (1998) 111--131.

\bibitem{Lau2006}
  Lap Chi Lau,
  Bipartite roots of graphs,
  {\sl ACM Transactions on Algorithms} 2 (2006) 178--208.

\bibitem{LauCor2004}
  Lap Chi Lau, Derek G.~Corneil,
  Recognizing powers of proper interval, split and chordal graphs,
  {\sl SIAM J. Discrete Math.} 18 (2004) 83--102.

\bibitem{LinSki1995}
  Yaw.-Ling Lin, Steven S.~Skiena,
  Algorithms for square roots of graphs,
  {\sl SIAM J. Discrete Math.} 8 (1995) 99--118.

\bibitem{Lubiw1982}
  A. Lubiw,
  $\Gamma$-free matrices,
  {\sl Master Thesis}, Dept. of Combinatorics and Optimization, University of Waterloo, Canada, 1982.

\bibitem{MotSud1994}
  Rajeev Motwani, Madhu Sudan,
  Computing roots of graphs is hard,
  {\sl Discrete Appl. Math.} 54 (1994) 81-88.

\bibitem{Muk1967}
  A. Mukhopadhyay,
  The square root of a graph,
  {\sl J. Combin. Theory} 2 (1967) 290-295.

\bibitem{Raych1992}
  A. Raychaudhuri,
  On powers of strongly chordal and circular arc graphs,
  {\sl Ars Combin.} 34 (1992) 147--160.

\bibitem{RosHar1960}
  I.C. Ross, Frank Harary,
  The square of a tree,
  {\sl Bell System Tech. J.} 39 (1960) 641--647.

\bibitem{TIAS}
  Shuji Tsukiyama, Mikio Ide, Hiromu Ariyoshi, and Isao Shirakawa,
  A new algorithm for generating all the maximal independent sets,
  {\sl SIAM J. Computing} 6 (1977) 505--517.

\end{thebibliography}
\end{document}